%% file: 01main_matching_function.tex
\documentclass[12pt]{article}
\usepackage[utf8]{inputenc}
\usepackage{amsmath,setspace,geometry}
\usepackage{amsthm}
\usepackage{amsfonts}
\usepackage[shortlabels]{enumitem}
\usepackage{rotating}
\usepackage{pdflscape}
\usepackage{graphicx}
\usepackage{bbm}
\usepackage[dvipsnames]{xcolor}
\usepackage{hyperref}
\hypersetup{colorlinks=true, linkcolor= BrickRed, citecolor = BrickRed, filecolor = BrickRed, urlcolor = BrickRed, hypertexnames = true}
\usepackage[]{natbib} 
\bibpunct[:]{(}{)}{,}{a}{}{,}
\usepackage[english]{babel}
\usepackage{float}
\usepackage{caption}
\usepackage{subcaption}
\usepackage{booktabs}
\usepackage{pdfpages}
\usepackage{threeparttable}
\usepackage{lscape}
\usepackage{bm}
\bibpunct[:]{(}{)}{,}{a}{}{,}
\usepackage{setspace}
\date{
This version: \today\\
First version: Oct 22, 2024
}
\begin{document}
\title{Nonparametric Estimation of Matching Efficiency and Elasticity on a Private On-the-Job Search Platform: Evidence from Japan, 2014-2024}
\author{Suguru Otani\thanks{\href{mailto:}{suguru.otani@e.u-tokyo.ac.jp}, Market Design Center, Department of Economics, University of Tokyo}\thanks{I thank Fuhito Kojima, Kosuke Uetake, Akira Matsushita, Kazuhiro Teramoto, Ryo Kambayashi, Higashi Yudai for their valuable advice. 
I thank Ken Edahiro, Shogo Hayashi, and Hayato Sasaki for sharing the data and their technical and institutional knowledge in the BizReach platform.
I thank Shunsuke Ishii for his excellent research assistance.
This work was supported by JST ERATO Grant Number JPMJER2301 and JSPS Grant-in-Aid (KAKENHI) for Young Researcher 25K16620, Japan.}
}
\maketitle

\begin{abstract}
\noindent

I analyze proprietary data from BizReach (2014–2024) to estimate the matching function for high-skill workers on a private on-the-job search platform using \cite{lange2020beyond} nonparametric approach. Comparing it to Hello Work, I find that matching efficiency on the private platform is both more volatile and higher, reflecting its growing popularity. Matching elasticity with respect to users is around 0.75, while for vacancies, it reaches 1.0, suggesting a more balanced elasticity than Hello Work. The study also uncovers industry-level heterogeneity, highlighting differences in matching dynamics across sectors.\\

\end{abstract}

\newpage

\section{Introduction}
On-the-job search plays a crucial role in labor reallocation, driving wage and productivity improvements \citep{moscarini2017relative}. In the U.S., job-to-job transitions account for one-third to one-half of all hires \citep{faberman2022job}. While Japan has historically been characterized by long-term employment stability, recent trends indicate a growing prevalence of on-the-job search. The proportion of employed individuals changing jobs remains relatively low at 4.8\% in 2023, but the share of workers actively seeking new opportunities reached a record high of 15.3\%, marking the tenth consecutive year of growth.\footnote{According to the Labor Force Statistics Office of the Statistics Bureau, Ministry of Internal Affairs and Communications.} Despite the increasing significance of on-the-job search in labor markets worldwide, empirical research on its matching efficiency and elasticity—especially in the context of private job platforms—remains scarce. This stands in contrast to the more extensive literature on job search behavior among unemployed individuals.

This paper fills this gap by examining long-term trends in on-the-job search platforms for high-skill workers across multiple labor markets. Using proprietary aggregate data from BizReach, a leading private online job scouting platform, I estimate the matching function and recover matching efficiency and elasticity through a nonparametric approach developed by \cite{lange2020beyond}. Unlike conventional job boards where workers actively apply for vacancies, BizReach allows registered job seekers to receive scouting messages from companies and headhunters searching for specialized talent. As of July 2024, over 2.58 million employed and self-employed workers were registered on the platform, underscoring its relevance in analyzing the dynamics of on-the-job search.

To contextualize these findings, I compare the private platform’s performance with the public employment service, Hello Work, using month-level aggregate data from the ``Report on Employment Service'' (\textit{Shokugyo Antei Gyomu Tokei}). This comparison provides broader insights into how private and public job search platforms function across different labor markets.

My results reveal key distinctions in matching efficiency and elasticity across platforms. For Hello Work, matching efficiency remained stable until 2021 before declining sharply. Elasticity with respect to unemployment was consistently low (0.1 to 0.4), whereas elasticity with respect to vacancies increased to 1.0, suggesting that vacancy fluctuations played a larger role in job matches than unemployment trends. In contrast, matching efficiency on BizReach was highly volatile, peaking in 2016, and exhibited generally higher elasticities than Hello Work. Elasticity with respect to users hovered around 0.75, while elasticity concerning vacancies rose steadily to 1.0 by 2024, indicating a more responsive matching process.

An industry-level analysis of the private platform highlights significant sectoral differences. The Consulting sector demonstrated higher matching efficiency and responsiveness to labor market fluctuations, particularly after 2020, whereas the IT and Manufacturing sectors showed more stable efficiency trends. These findings emphasize industry heterogeneity in labor market dynamics and suggest that private platforms function differently across sectors compared to public employment services.

By leveraging proprietary aggregate data, this study provides novel empirical insights into the matching function of on-the-job search labor markets beyond a single national context. While the BizReach platform serves as a valuable case study, the findings contribute to a broader understanding of private job market dynamics and their role in labor reallocation across different economies.

\subsection{Related literature}
This paper contributes to three key areas of research: nonparametric matching functions, on-the-job search, and online job search platforms operated by private firms.

First, it adds to the empirical literature on the estimation of the matching function, a foundational component in macroeconomic models. Using a novel nonparametric approach developed by \cite{lange2020beyond}, I examine trends in matching efficiency in Japanese labor markets via an online job scouting platform. This method enables the identification and estimation of the matching function without imposing the standard independence assumption between matching efficiency and search efforts from either side of the labor market. This approach accommodates multiple types of job seekers. \cite{lange2020beyond} highlight the positive correlation between efficiency and market structure variables like labor market tightness, which introduces a positive bias in vacancy elasticity estimates unless unobserved matching efficacy is accounted for. In traditional Cobb-Douglas matching function models, this unobserved factor is often ignored, resulting in potentially biased elasticity estimates.\footnote{For example, \cite{petrongolo2001looking} summarize early aggregate studies using the Cobb-Douglas matching function, finding the match elasticity with respect to unemployment to be in the range of 0.5 to 0.7. In the context of Japan, \cite{otani2024nonparametric} update the earlier findings of \cite{kano2005estimating}, \cite{kambayashi2006vacancy}, and others, comparing results with international findings like \cite{bernstein2022matching}.}

This paper also relates to the Japanese labor market literature, particularly in the context of the public off-the-job search platform, Hello Work.\footnote{See the survey of \cite{miyamoto2025macroeconomic} for an overview of labor markets in Japan.} \cite{otani2024nonparametric} estimate matching efficiency and mismatch in Japan's Hello Work platform between 1972 and 2024, showing a declining trend in matching efficiency consistent with decreasing job and worker finding rates. The match elasticity with respect to unemployment is found to be between 0.5 and 0.9, while the elasticity concerning vacancies ranges from 0.1 to 0.4. In comparison, \cite{kanayama2024nonparametric} apply a similar nonparametric method to estimate matching efficiency and elasticity in a privately-operated spot-worker platform, contrasting these findings with Hello Work's part-time data. My paper complements and extends these findings by focusing on a high-skill, employed worker platform, filling an important gap in the literature.


The Japanese government’s efforts to enhance labor market flexibility have significantly contributed to the increase in on-the-job search activities. A prominent example is the 2018 Work Style Reform Acts, which aimed to reduce overwork, improve work-life balance, and promote diverse career trajectories. While these reforms were not directly designed to foster job transitions, measures such as the promotion of side jobs and secondary employment have encouraged workers to explore new career pathways, helping to reduce the societal stigma previously associated with job changes.

At the same time, the expansion of educational training benefits through the Employment Insurance System has provided workers with greater opportunities to reskill and acquire advanced competencies. Together with structural pressures from persistent labor shortages since the 2010s, these developments have further reshaped attitudes toward mobility and created an environment more conducive to job transitions. The growth of the job transition market should therefore be understood not as the product of the Work Style Reform Acts alone, but rather as the outcome of a broader set of institutional reforms and structural changes. This policy context underscores the relevance of quantitative studies on private on-the-job search platforms, such as those examined in this paper.

Second, this paper contributes to the literature on on-the-job search, a key aspect of labor search theory since the 1970s \citep{burdett1978theory}. Recent theoretical models, including those by \cite{cahuc2006wage}, \cite{eeckhout2019unemployment}, and \cite{bagger2019empirical}, emphasize the role of search effort in on-the-job search and its connection to job ladder dynamics. Empirical research, such as \cite{mueller2010job} and \cite{ahn2017precautionary}, has relied on data from the American Time Use Survey (ATUS) to document on-the-job search behaviors. However, due to the limitations of ATUS in capturing search outcomes, there is a gap in evaluating the efficiency of on-the-job search. Notably, \cite{faberman2022job} and \cite{roussille2023bidding} provide crucial insights, with \cite{faberman2022job} focusing on the relationship between search effort and outcomes and \cite{roussille2023bidding} exploring wage markdown using data from Hired.com. However, these studies lack long-term macro-level insights into matching function, efficiency, and elasticity. The proprietary data in this paper offers a unique advantage by allowing for the evaluation of matching efficiency in the on-the-job search labor market via a private platform.

\begin{figure}[!ht]
  \begin{center}
  \includegraphics[width = 0.77\textwidth]
  {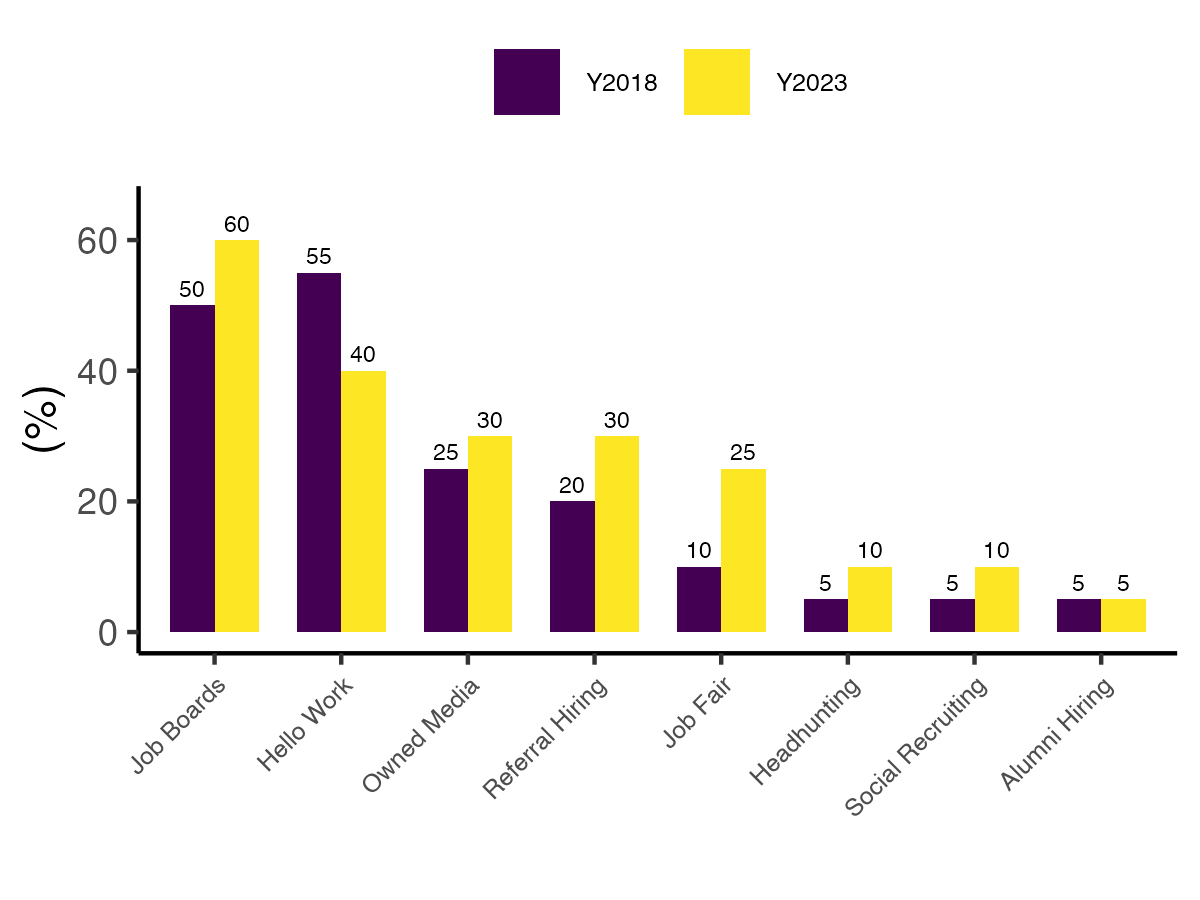}
  \caption{Mid-Career Recruitment Channels Prioritized by Companies}
  \label{fg:recruitement_channel} 
  \end{center}
  \footnotesize
  Source: Cabinet Office, Government of Japan (2024). The author reproduces Figure 2 in ``2024 Annual Economic and Fiscal Report" Chapter 2. The survey asks sampled firms whether they evaluate each channel or not.
\end{figure}

Third, this paper contributes to the expanding literature on online job search platforms. The analysis of job matching within real-world market institutions has gained prominence due to the increasing availability of data from online job platforms \citep{autor2019studies}.\footnote{Examples include studies like \cite{kuhn2004internet}, \cite{kuhn2014internet}, and \cite{kroft2014does}, which focus on worker status, while others, such as \cite{kuhn2013gender}, \cite{hershbein2018recessions}, \cite{brown2016boarding}, and \cite{azar2020concentration}, focus solely on vacancy data. Moreover, research like \cite{banfi2019high}, \cite{marinescu2018mismatch}, \cite{marinescu2020opening}, and \cite{azar2022estimating} incorporates both worker and vacancy information.} Much of the literature emphasizes application-level or vacancy-level behavior to assess search behavior and wage elasticity. For instance, \cite{faberman2019intensity} leverage proprietary application-level data from an online job search engine to explore the relationship between search intensity and duration, primarily focusing on lower-skill, hourly jobs for employed and unemployed workers. Similarly, \cite{kambayashi2025decomposing} estimate elasticities of application, interview attendance, and offer acceptance relative to posted wages using detailed process-level data from private job search and matching intermediary platforms in Japan, which became the most significant recruitment channel in 2023, as shown in Figure \ref{fg:recruitement_channel}. In contrast, this paper adopts a broader macro-level perspective, evaluating the overall efficiency of the private matching platform. To the best of my knowledge, this is the first paper to estimate matching efficiency and elasticity in an online job scouting platform, offering relatively long-term insights into private online job search trends, complementing the micro-level studies.

\section{Data}

\subsection{Data source}

First, I use the Report on Employment Service (\textit{Shokugyo Antei Gyomu Tokei}) for month-level aggregate data from January 2014 to April 2024 to examine trends in matching unemployed workers with vacancies via Japan's public employment platform, Hello Work. These datasets include the number of job openings, job seekers, and successful job placements, primarily sourced from the Ministry of Health, Labour and Welfare (MHLW) of Japan, which regularly publishes monthly reports and statistical data on the Public Employment Security Office, commonly known as Hello Work. Hello Work plays a crucial role in Japan's labor market by providing government-operated employment counseling, job placement services, and vocational training. It has been extensively used for estimating traditional Cobb-Douglas matching functions, as seen in studies like \cite{kano2005estimating}, \cite{kambayashi2006vacancy}, \cite{sasaki2008matching}, and \cite{higashi2018spatial}, as well as nonparametric estimation \cite{otani2024nonparametric}. In this study, I focus on full-time workers to ensure consistency for comparison across different datasets. The chosen period provides a consistent timeframe for comparison with the following platform data.

Second, I utilize proprietary data from BizReach, a private job-scouting platform in Japan, to analyze trends in matching employed workers with vacancies. To maintain consistency with the Hello Work data, I include only ``active" workers, defined as those who logged into the platform in a given month, excluding inactive registered users. Unlike Hello Work, BizReach caters to high-level professionals and executives, offering a premium job-scouting service. Candidates can either use the platform for free or pay a monthly subscription fee (approximately 40 U.S. dollars) to gain priority access to job opportunities and services.

The BizReach platform allows job seekers to upload resumes and receive scouting messages from companies or headhunters searching for specialized talent. This system encourages proactive recruitment, enabling direct communication between job seekers and employers. Users also gain insights into their market value through the scouts they receive, even if they are not actively searching for new opportunities. This platform’s focus on high-skill professionals contrasts with the broader services provided by Hello Work, which includes support for entry-level and part-time positions. While BizReach emphasizes efficiency in high-level recruitment, it may not be suitable for individuals seeking entry-level roles or more comprehensive career counseling services, which Hello Work provides.

\subsection{Trend comparison}

\begin{figure}[!ht]
  \begin{center}
  \subfloat[User $U$, vacancy $V$, and tightness ($\frac{V}{U}$)]{\includegraphics[width = 0.37\textwidth]
  {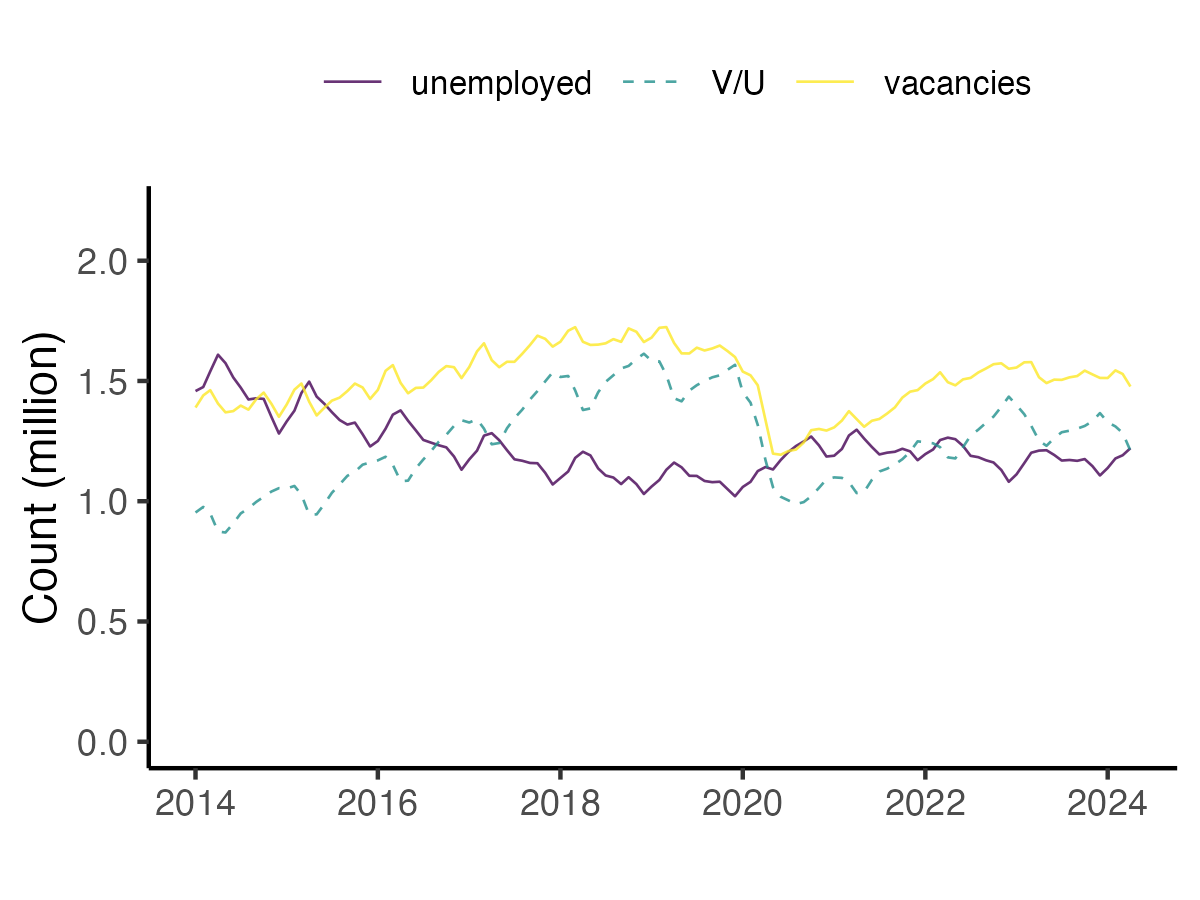}\includegraphics[width = 0.37\textwidth]
  {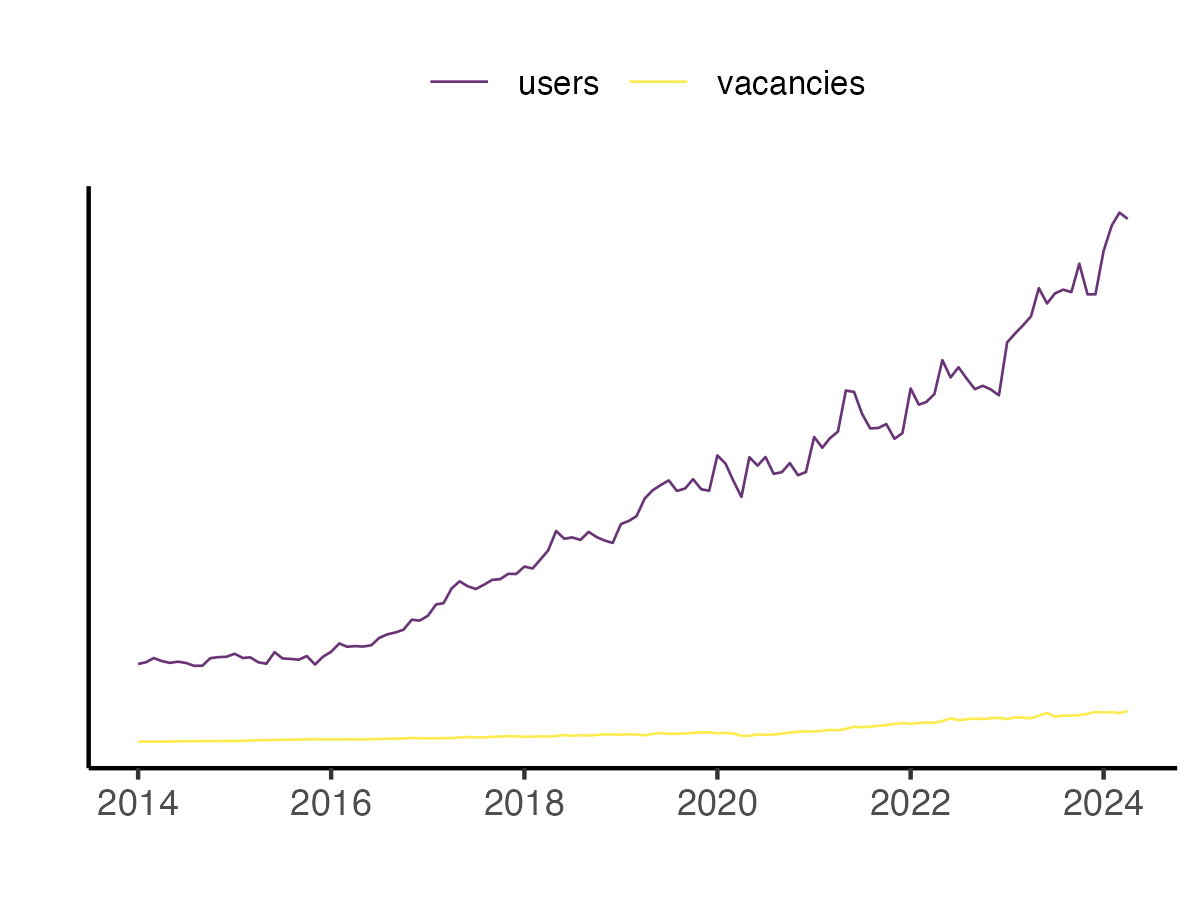}}\\
  \subfloat[Hire $H$]{\includegraphics[width = 0.37\textwidth]
  {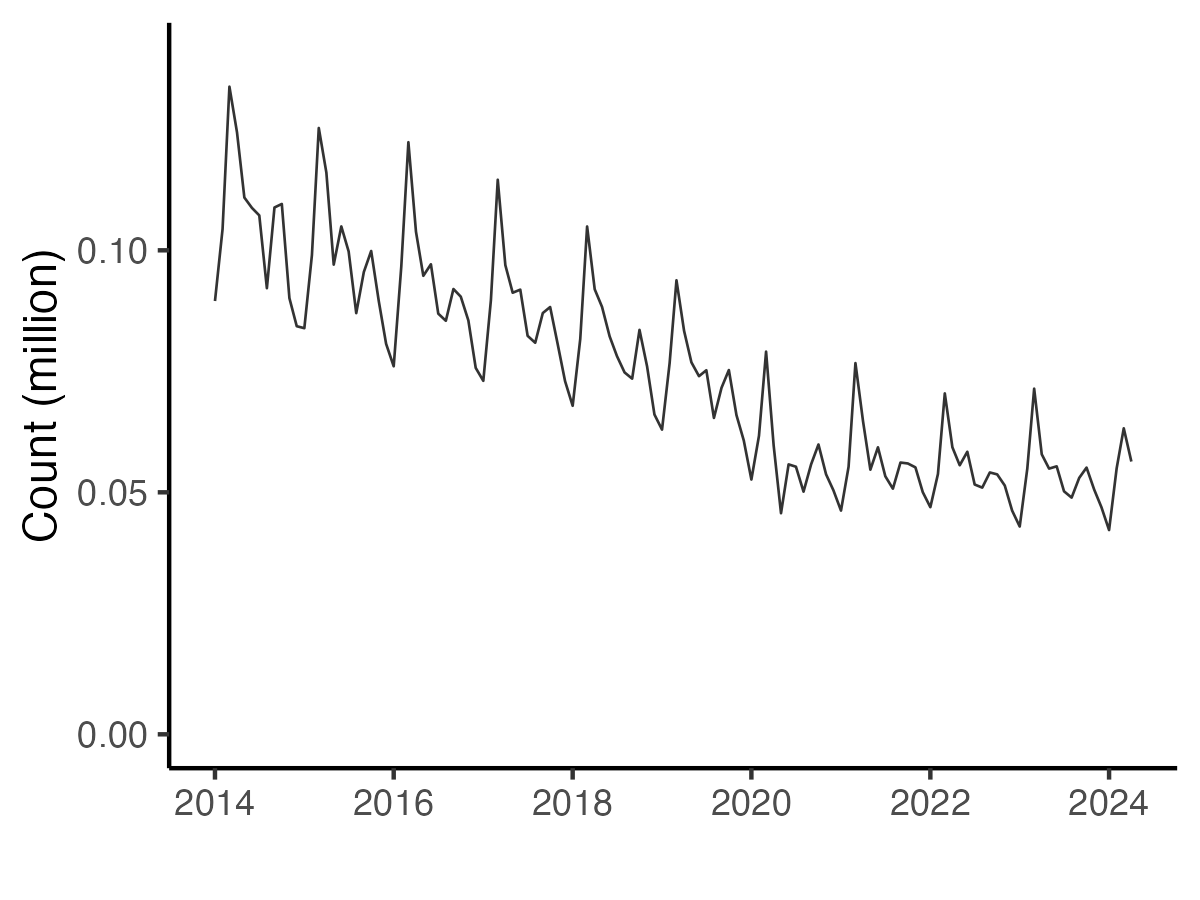}\includegraphics[width = 0.37\textwidth]
  {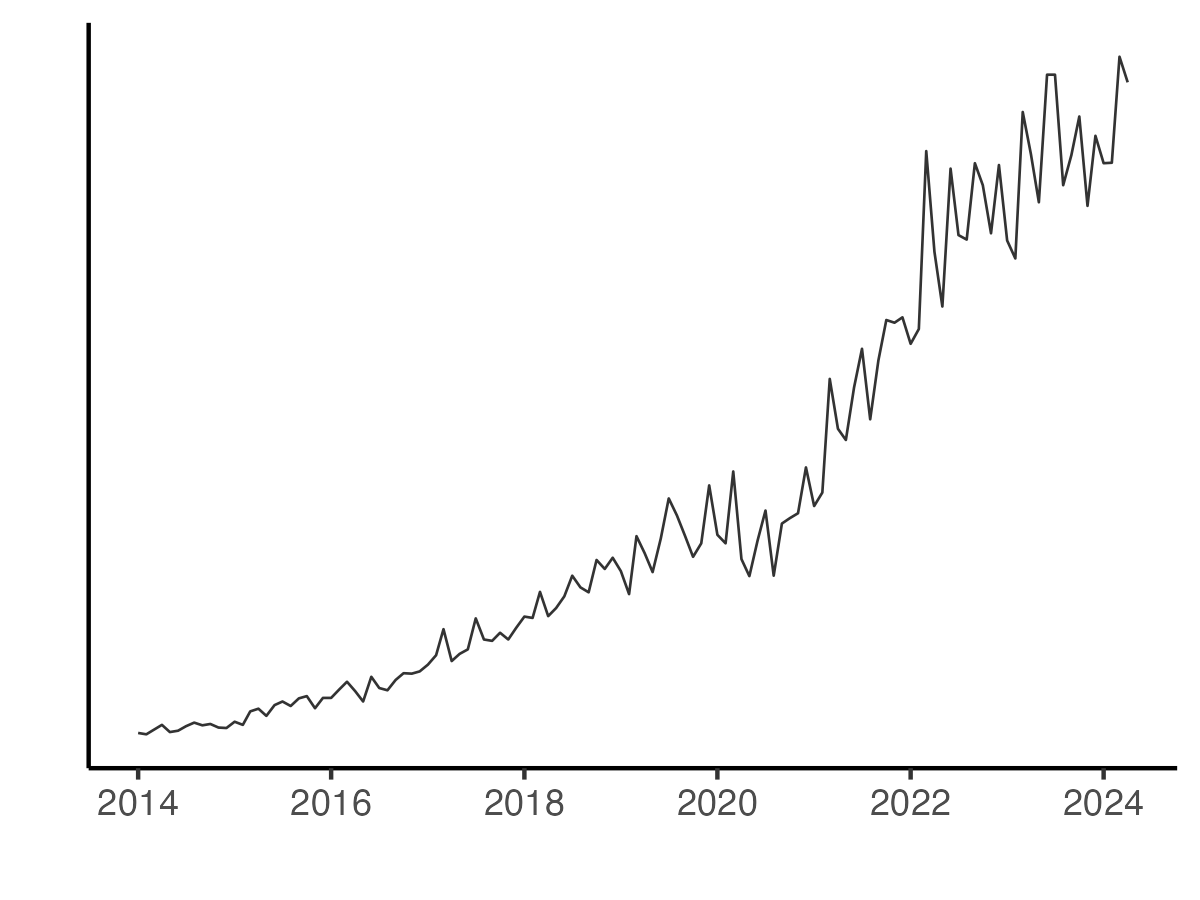}}
  \\
  \subfloat[Job Worker finding rate ($\frac{H}{U}$,$\frac{H}{V}$)]{\includegraphics[width = 0.37\textwidth]
  {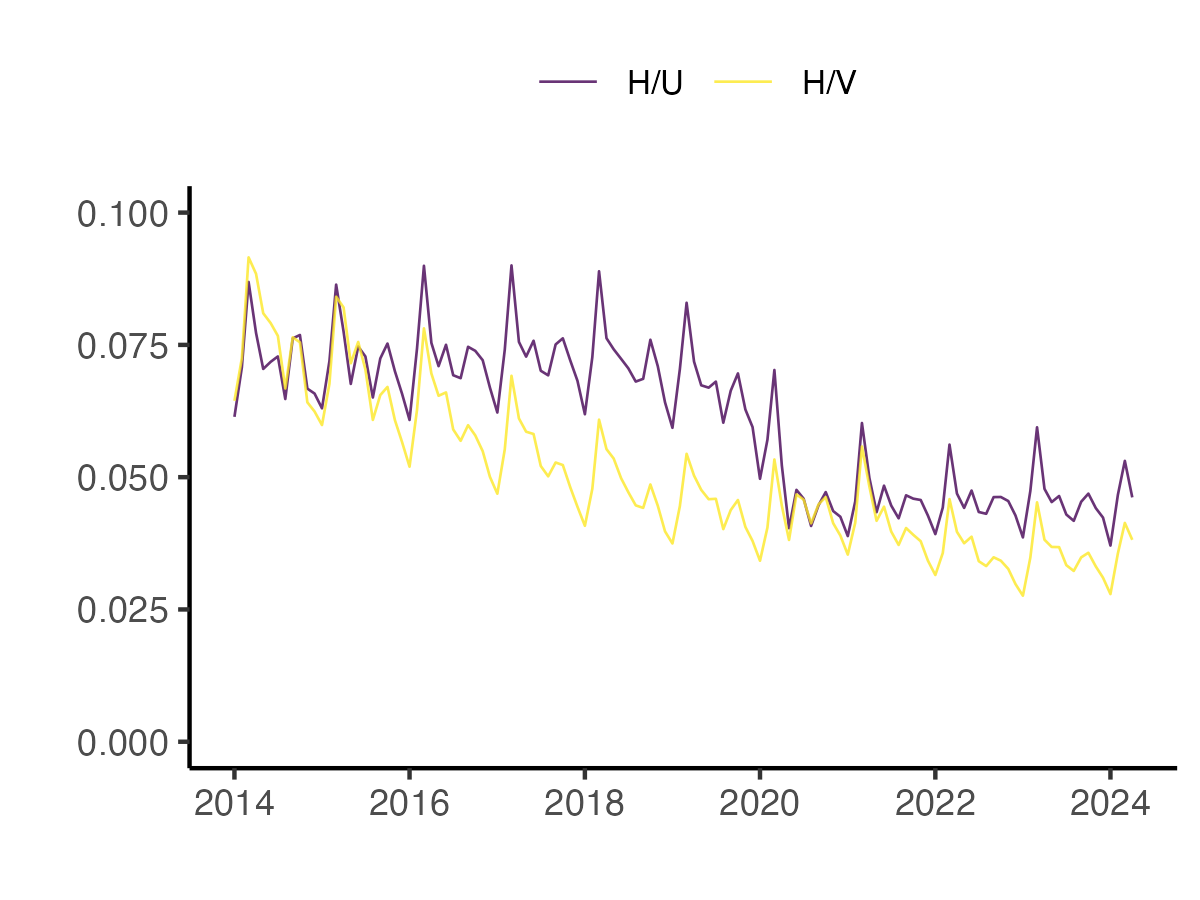}\includegraphics[width = 0.37\textwidth]
  {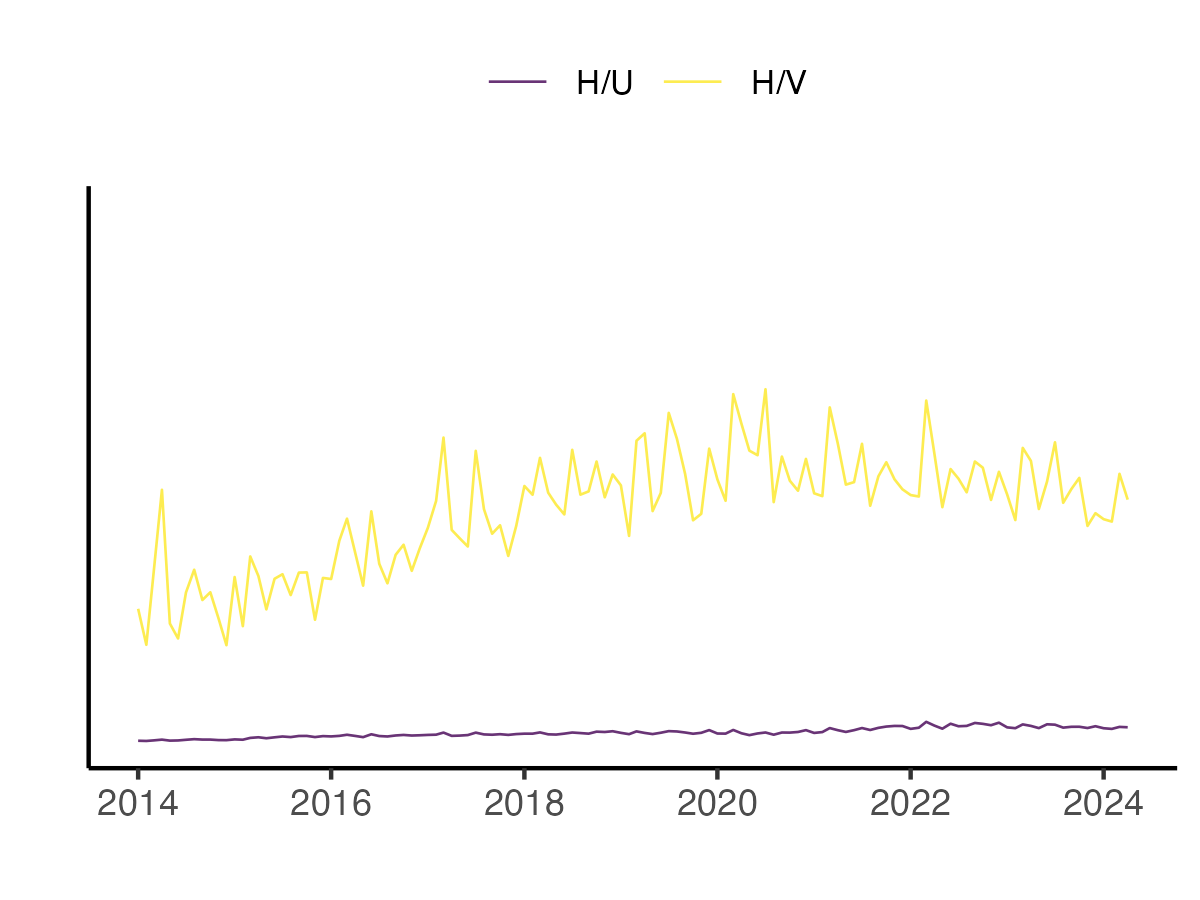}}
  \caption{Trends of key variables: Hello Work full-time (left) vs platform (right) 2014-2024}
  \label{fg:unemployed_vacancy_month_aggregate} 
  \end{center}
  \footnotesize
  Note: For confidentiality reasons, the y-axis levels in the right panels have been masked, making them not directly comparable with those in the left panels. Additionally, labor market tightness on the platform has not been reported to maintain confidentiality.
\end{figure} 

Figure \ref{fg:unemployed_vacancy_month_aggregate} provides a comparative analysis of labor market dynamics between the Hello Work public employment platform (left panel) and a private scouting platform (right panel) from 2014 to 2024. For confidentiality reasons, the y-axis levels for the platform panels have been masked and are therefore not directly comparable with Hello Work panels. 

In the Hello Work panel, the unemployment trend remains relatively stable, fluctuating around 1.5 million over the observed period, with a slight decline noted after 2020. The number of vacancies increases steadily, resulting in a moderate rise in labor market tightness, although this ratio stays below 1 throughout the period. Both the hiring count and the job-worker finding rates which are the numbers of matching hires divided by the number of vacancies and workers exhibit a downward trend, suggesting two possible explanations. The first is the existence of potential challenges, inefficiencies, or mismatches in job placements facilitated by the Hello Work platform. The second alternative explanation could be the growing presence of private job search platforms, which offer an alternative avenue for unemployed workers seeking career transitions. It is important to note, however, that the private platform analyzed in this paper primarily caters to employed workers, thus limiting direct evidence regarding its impact on unemployed workers.

The right panels in the updated figure illustrate labor market trends on the private platform for high-skilled employed workers from 2014 to 2024. There is a steady and consistent increase in employed users, especially starting around 2018. According to the Japanese Labor Force Survey, the total number of employed workers is 66 million in 2014 and 69 million in 2024. Additionally, the Labor Force Statistics Office reports that approximately 10.35 million employed individuals were seeking to change jobs or explore new opportunities in 2023. Thus, around 1\% of the total employed workforce and nearly 5\% of those actively exploring job changes have registered on the platform, indicating their engagement in on-the-job search activities. However, this does not necessarily imply active job pursuit for all registered users. Conversely, vacancies increase gradually but remain relatively modest, resulting in a low labor market tightness ratio, which is not reported above for confidentiality reasons. This low tightness ratio does not necessarily indicate a mismatch between job supply and demand on the platform, as registered workers may be passively exploring opportunities rather than actively searching for new positions.

The right panel of Figure (b) illustrates the hiring count, which shows a gradual and steady increase beginning around 2018. However, even with this upward trend, the overall hiring level masked in the figure remains relatively modest compared to Hello Work, particularly when considering the substantial rise in the number of users. This pattern may suggest that many users are engaging with the platform in a more passive manner rather than actively seeking new employment. Panel (c) presents the job-worker finding rates, with the worker-finding rate displaying slight fluctuations but maintaining a generally comparable level to that of Hello Work. Meanwhile, the job-finding rate remains consistently low throughout the period compared to the worker-finding rate. These patterns suggest a notable disparity between worker-side and vacancy-side matching probabilities, despite the marked growth in platform users and job postings.

\section{Model}
Our primary focus lies in analyzing matching efficiency and matching elasticity with respect to the number of registered workers and vacancies in the labor market, as facilitated by an online job scouting platform operated by a private firm in Japan. A matching function derived from search models plays a pivotal role in labor economics.\footnote{See \cite{pissarides2000equilibrium}, \cite{petrongolo2001looking}, and \cite{rogerson2005search} for further reference.} The matching function operates on the premise of random search from both sides of the labor market, where job seekers represent labor supply and recruiters represent labor demand.
Conceptually, this paper examines two independent matching functions on the off-the-job and on-the-job search labor markets.
Examining interdependence between two matching functions theoretically is out of the scope of this paper.

Note that our analysis does not aim to equate the two markets, but rather to compare system-level matching efficiency patterns within a consistent empirical framework. While the matching function itself is agnostic to the interpretation of matching efficiency, the implications differ across platforms. In the case of Hello Work, matching efficiency reflects the search effort of unemployed, typically lower-skilled workers seeking full-time jobs, or the effectiveness of Hello Work’s employment support services. In contrast, matching efficiency on the BizReach platform reflects the passive search effort of employed high-skilled workers awaiting scouting offers—driven primarily by recruiters’ hiring initiatives—or the effectiveness of the platform’s services, including its user interface and recommendation algorithms.

To estimate the matching function and recover matching efficiency, I adopt the novel approach proposed by \cite{lange2020beyond}. The paper highlights two critical issues: the endogeneity of matching efficiency \citep{borowczyk2013accounting} and the overly restrictive nature of the Cobb-Douglas specification, which assumes fixed matching elasticity. To address these limitations, \cite{lange2020beyond} propose a nonparametric identification and estimation framework for matching efficiency under specific conditions that will be discussed later.

Let unscripted capital letters $(A, U, V)$ denote random variables, while time-specific realizations are subscripted by $t$. I consider the matching function $m_t(\cdot,\cdot)$, which maps period-$t$ users $U_t$, per-capita search efficiency/matching efficiency $A_t$, and vacancies $V_t$ into hires $H_t$. For the private platform analysis, $U_{t}$ represents the number of employed workers registered on the platform, whereas $U_{t}$ for Hello Work refers to unemployed workers registered in the public system. I assume a stationary data-generating process, with sufficient time-series data to treat the joint distribution $G: \mathbb{R}_{+}^3 \rightarrow[0,1]$ of $\left(H_t, U_t, V_t\right)$ as observable. Additionally, I denote by $F(A, U)$ the joint distribution of $A$ and $U$.

I identify the matching function and the unobserved, time-varying matching efficiency, $A$. First, I assume that $V$ and $A$ are conditionally independent given $U$, i.e., $A \perp V \mid U$.\footnote{Matching models often incorporate a free entry condition, whereby firms continue to post job vacancies as long as the expected return is positive. This condition implies that the number of vacancies is endogenously determined and may depend on current matching efficiency, potentially violating the exogeneity assumption \citep{borowczyk2013accounting}. While \cite{lange2020beyond} additionally incorporate search effort data \citep{mukoyama2018job} and a recruitment intensity index data \citep{davis2013establishment} to attempt to address this endogeneity issue using time-use survey data—which is typically unavailable in most contexts—\cite{otani2024nonparametric} instead demonstrates through Monte Carlo simulations that the resulting bias is economically negligible, with estimated parameters deviating by less than 5\% from the time-varying true values, and benefit of capturing time-varying efficiency is substantial.} 
Second, I assume that the matching function $m(AU, V): \mathbb{R}_{+}^2 \rightarrow \mathbb{R}$ exhibits constant returns to scale (CRS).\footnote{To align with the original model of \cite{matzkin2003nonparametric}, the function $H=m(AU,V)$ can be reformulated as $H/U=m(A,V/U)$ under CRS, where $H/U$ and $V/U$ are the job-finding rate and market tightness, respectively.} These two assumptions are commonly used in the literature. Applying the nonparametric identification results of \cite{matzkin2003nonparametric}, Proposition 1 of \cite{lange2020beyond} demonstrates that the joint distribution $G(H, U, V)$ identifies $F(A, U)$ and the matching function $m(AU, V): \mathbb{R}_{+}^2 \rightarrow \mathbb{R}_{+}$ up to a normalization of $A$ at one point, denoted as $A_0$, within the support of $(A, U, V)$.\footnote{In \cite{otani2024nonparametric}, I report finite sample performance and extend the methodology through Monte Carlo simulation. Simulation results with a sample size of $T=50$ indicate that the sample size in this paper is sufficient for accurately recovering matching efficiency. For practical issues and application in different contexts, see \cite{brancaccio2020guide}, which applies this approach to estimate a matching function in a trade model \citep{brancaccio2020geography,brancaccio2023search}, and \cite{otani2025nonparametric} which applies this approach to marriage markets.}

\section{Estimation}

Following \cite{lange2020beyond}, I begin by estimating \( F(A_0 | U) \) across the support of \( U \). To achieve this, I use the distribution of hires conditional on users, \( U \), and observed vacancies, \( V \). Specifically, we have:
\begin{align*}
    F(A_0|\psi U_0) = G_{H|U,V}(\psi H_0|\psi U_0, \psi V_0) \quad \text{for any arbitrary scalar } \psi,\\
    F(\psi A_0|\lambda U_0) = G_{H|U,V}(\psi H_0|\lambda U_0, \psi V_0) \quad \text{where } \lambda > 0 \text{ is a scaling factor},
\end{align*}
where \( F(A_0|\psi U_0) \) and \( G_{H|U,V} \) represent the respective conditional distributions. By varying the parameters \( (\psi, \lambda) \), I can trace out \( F(A | U) \) across the entire support of \( (A, U) \).

Given that my data is finite, I rely on an estimate of \( G_{H|U,V} \) for the constructive estimator. Consider an arbitrary point \( (H_\tau, U_\tau, V_\tau) \). To obtain \( G(H_\tau | U_\tau, V_\tau) \), I calculate the proportion of observations with fewer hires than \( H_\tau \), taken from observations proximate to \( (U_\tau, V_\tau) \) in the \( (U, V) \)-space. In practice, this is done by averaging across all observations, assigning smaller weights to those with values \( (U_t, V_t) \) distant from \( (U_\tau, V_\tau) \) via a kernel that discounts distant observations. The resulting estimate of $F(\psi A_0 | \lambda U_0) = G_{H|U,V}(\psi H_0 | \lambda U_0, \psi V_0)$ is expressed as:
\begin{align*}
    \hat{F}(\psi A_0 | \lambda U_0) = \sum 1(H_t < \psi H_0) \kappa(U_t, V_t, \lambda U_0, \psi V_0),
\end{align*}
where \( \kappa(.) \) denotes a bivariate normal kernel with bandwidth 0.01.

Once the distribution function \( F(A | U) \) is recovered, I invert \( F(A_t | U_t) \) to derive \( A_t \) for all observations in the dataset, using:
\begin{align*}
    A_t = F^{-1}(G(H_t | U_t, V_t) | U_t),
\end{align*}

Finally, I recover the matching function as:
\begin{align*}
    m(A_t U_t, V_t) = G^{-1}(F(A_t | U_t) | U_t).
\end{align*}

To compute matching elasticities, I employ a LASSO regression, projecting hires onto the original and squared values of vacancies and users, interacted with implied matching efficiency. The resulting estimates approximate the derivatives of the matching function with respect to vacancies and users, interacted with implied matching efficiency. This provides an estimate of the elasticity of the matching function.\footnote{The matching elasticity with respect to users $\frac{d \log m(AU,V)}{d \log U}=\frac{d m(AU,V)}{d U}\frac{U}{H}=\frac{d m(AU,V)}{d AU}\frac{d AU}{dU}\frac{U}{H}=\frac{d m(AU,V)}{d AU}\frac{AU}{H}=\frac{d \log m(AU,V)}{d \log AU}$ is obtained from the regression coefficient of $H$ on $AU$ and multiplying it by $\frac{AU}{H}$. Concretely, I approximate $m$ by the second order polynomial $AU$ and $V$.}

\section{Results}

\subsection{Matching efficiency and elasticity in the platform}

\begin{figure}[!ht]
  \begin{center}
  \subfloat[Matching Efficiency ($A$)]{\includegraphics[width = 0.37\textwidth]
  {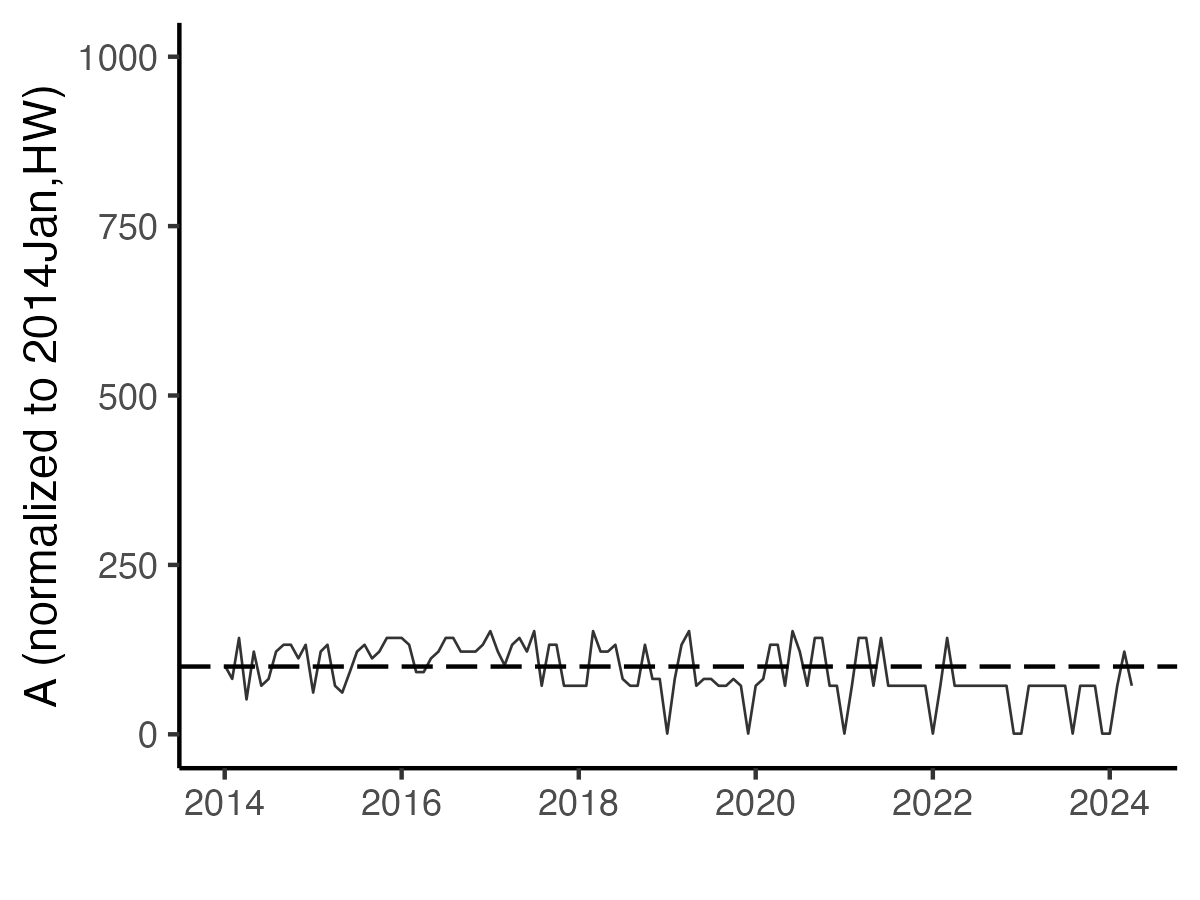}\includegraphics[width = 0.37\textwidth]
  {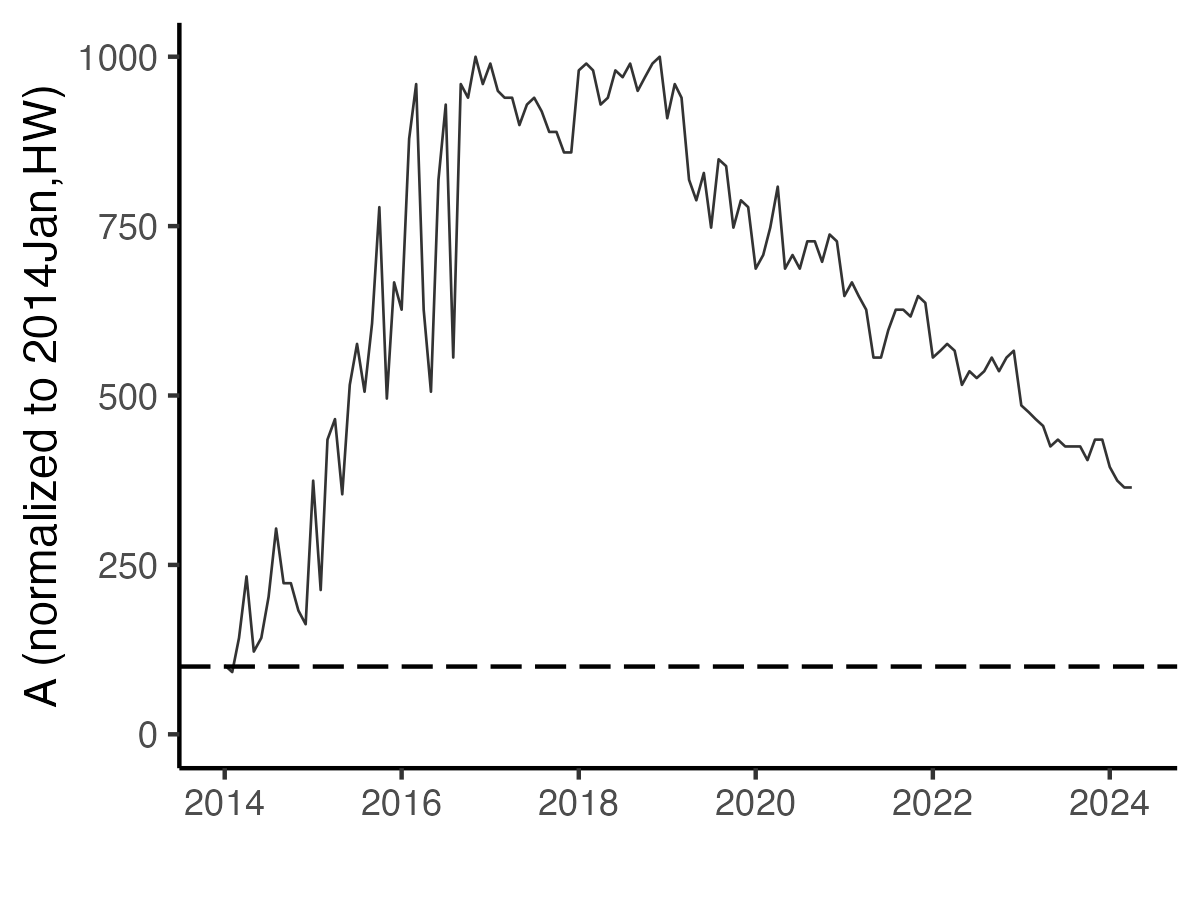}}\\
  \subfloat[Matching Elasticity ($\frac{d\ln m}{d \ln AU}$, $\frac{d\ln m}{d\ln V}$)]{\includegraphics[width = 0.37\textwidth]
  {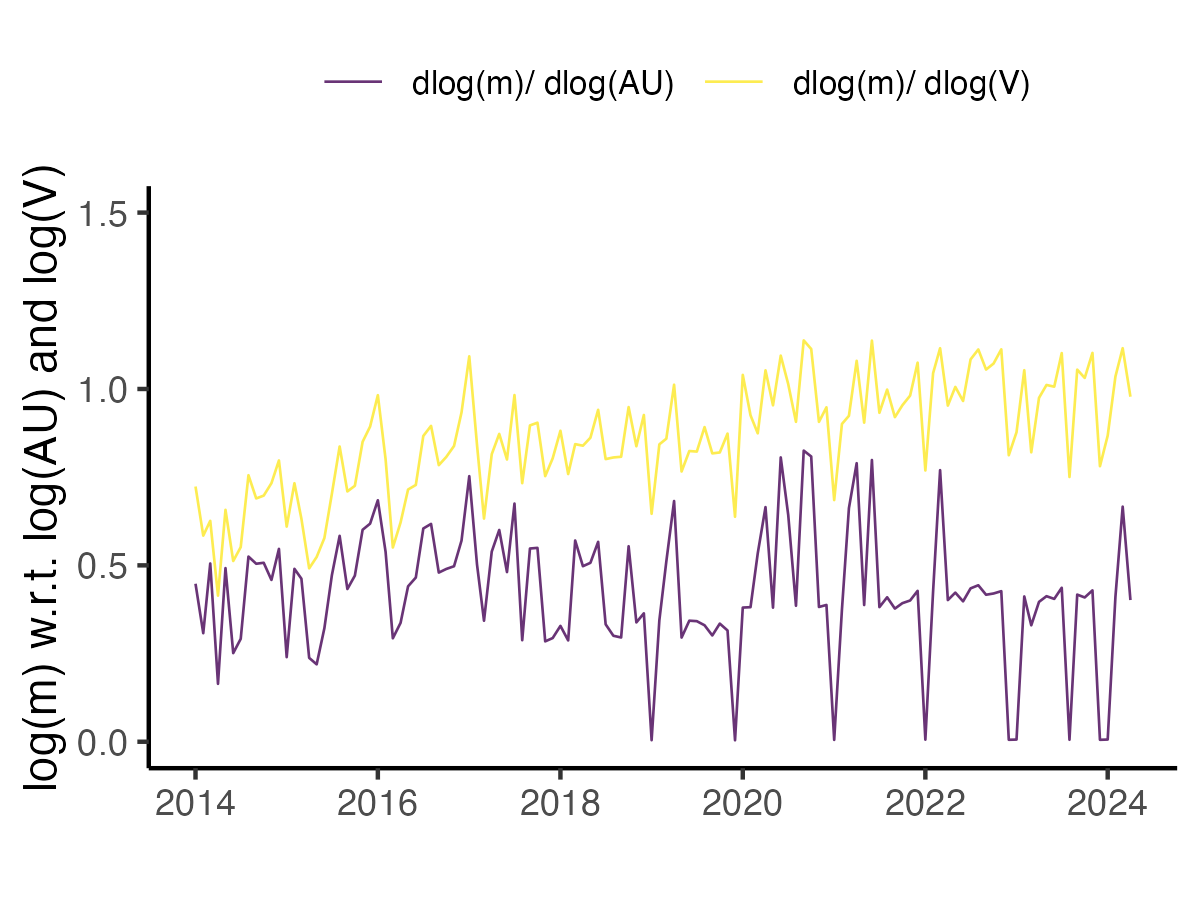}\includegraphics[width = 0.37\textwidth]
  {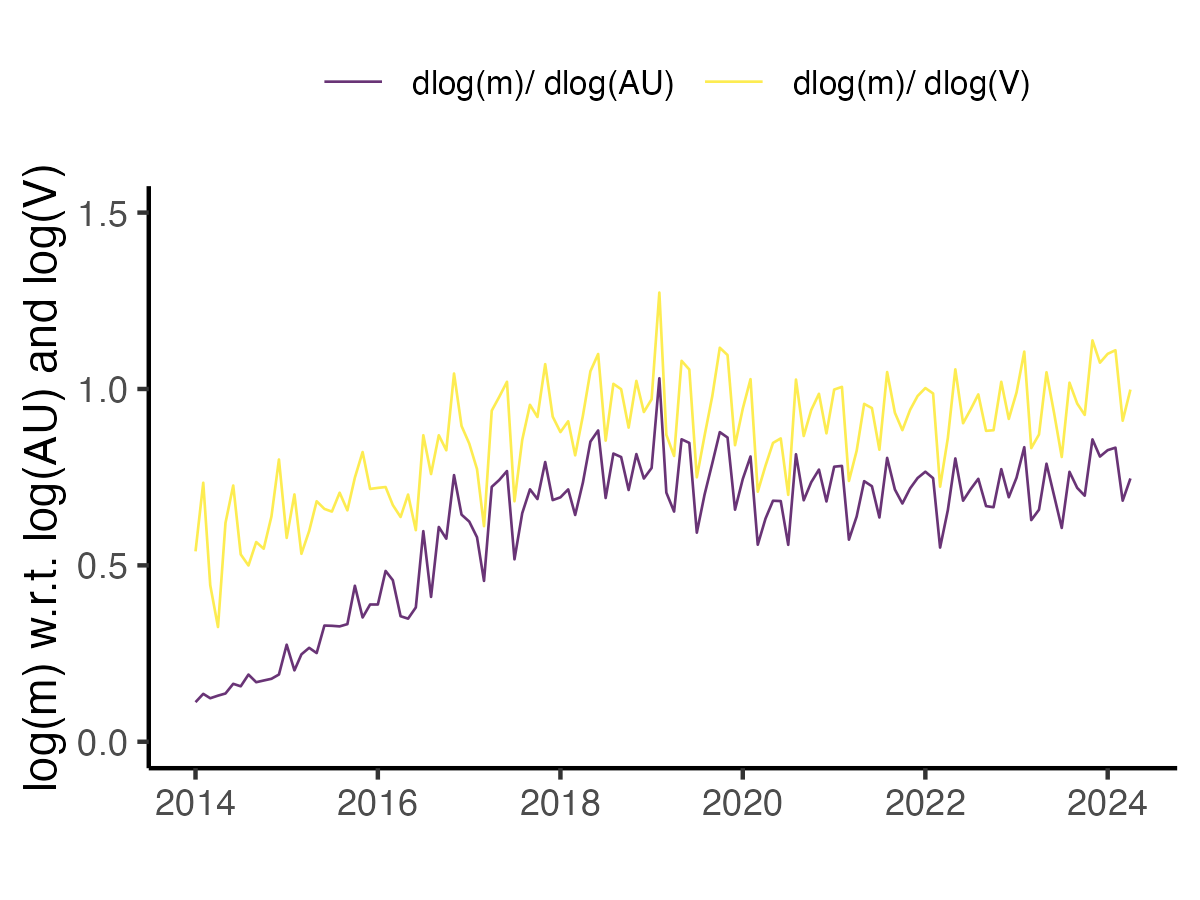}}
  \caption{Hello Work full-time vs platform 2014-2024}
  \label{fg:matching_efficiency_month_aggregate} 
  \end{center}
  \footnotesize
\end{figure} 

Figure \ref{fg:matching_efficiency_month_aggregate} presents a comparison of matching efficiency and matching elasticity between the Hello Work public employment platform (left panel) and a private scouting platform (right panel) from 2014 to 2024. I set the reference period to $t = 0$ corresponding to January 2014, and normalize matching efficiency by setting $A_0 = 100$. In the Hello Work platform, as depicted in panel (a), matching efficiency remains relatively stable around the baseline (normalized to January 2014) until 2021 but sharply declines thereafter. Panel (b) illustrates the matching elasticities with respect to unemployment and vacancies. The elasticity with respect to unemployment remains consistently lower, around 0.4, while the elasticity with respect to vacancies shows a gradual increase, reaching values near 1.0 by 2022. This suggests that changes in vacancies have a larger impact on the number of matches than changes in unemployment, and the assumption of constant returns to scale has been seemingly violated after 2020.\footnote{The estimated elasticities differ from those reported in \cite{otani2024nonparametric}, largely due to the different time horizons considered. Specifically, \cite{otani2024nonparametric} includes data from 1972 to 2024, a period that encompasses various economic booms and busts, which likely captures a broader range of labor market dynamics and affects the elasticity estimates. Also, see the Cobb-Douglas specification results in Appendix \ref{sec:appendix_cobb_douglas}. }

On the private scouting platform (right panel), panel (a) shows significantly higher volatility in matching efficiency, especially after 2016, where it peaks sharply above 900 before declining and stabilizing around 400 by 2024. This indicates that the private platform experienced large fluctuations in its ability to efficiently match job seekers with vacancies, potentially due to shifts in demand or changes in the platform's user base. Panel (b) displays the matching elasticities with respect to users and vacancies on the private platform. The elasticity with respect to users remains consistently around 0.75, while the elasticity with respect to vacancies is generally higher, rising steadily and reaching around 1.0 by 2024, and the assumption of constant returns to scale has been seemingly violated after 2020. This suggests a more responsive matching process to changes in the number of vacancies. The higher volatility and responsiveness of the private platform, along with a more balanced elasticity between users and vacancies, highlight a key distinction from Hello Work, where the dynamics are comparatively more stable but less responsive.

\subsection{Industry-level matching efficiency and elasticity in the platform}

\begin{figure}[!ht]
  \begin{center}
  \subfloat[Tightness ($V/U$)]{\includegraphics[width = 0.37\textwidth]  {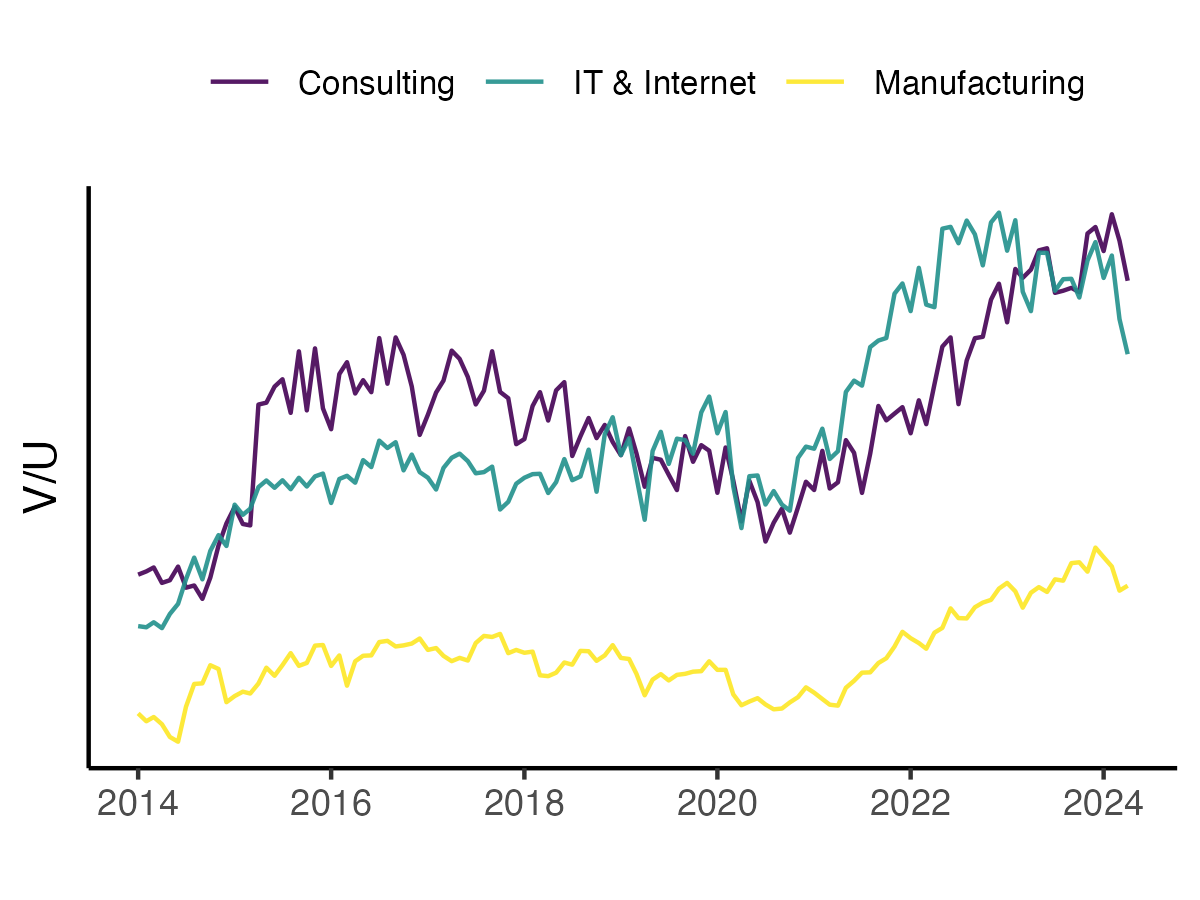}}
  \subfloat[Matching Efficiency ($A$)]{\includegraphics[width = 0.37\textwidth]
  {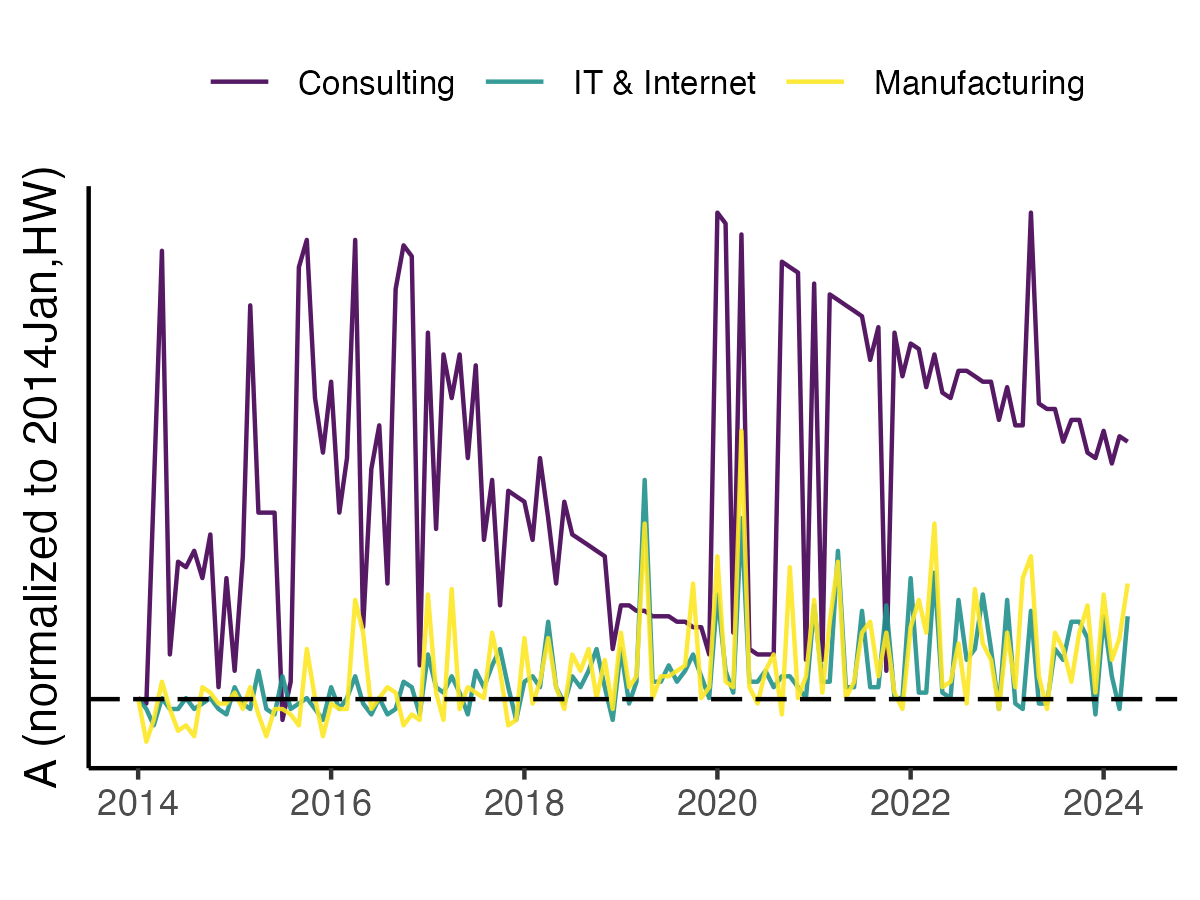}}\\
  \subfloat[Matching Elasticity ($\frac{d\ln m}{d \ln AU}$)]{\includegraphics[width = 0.37\textwidth]
  {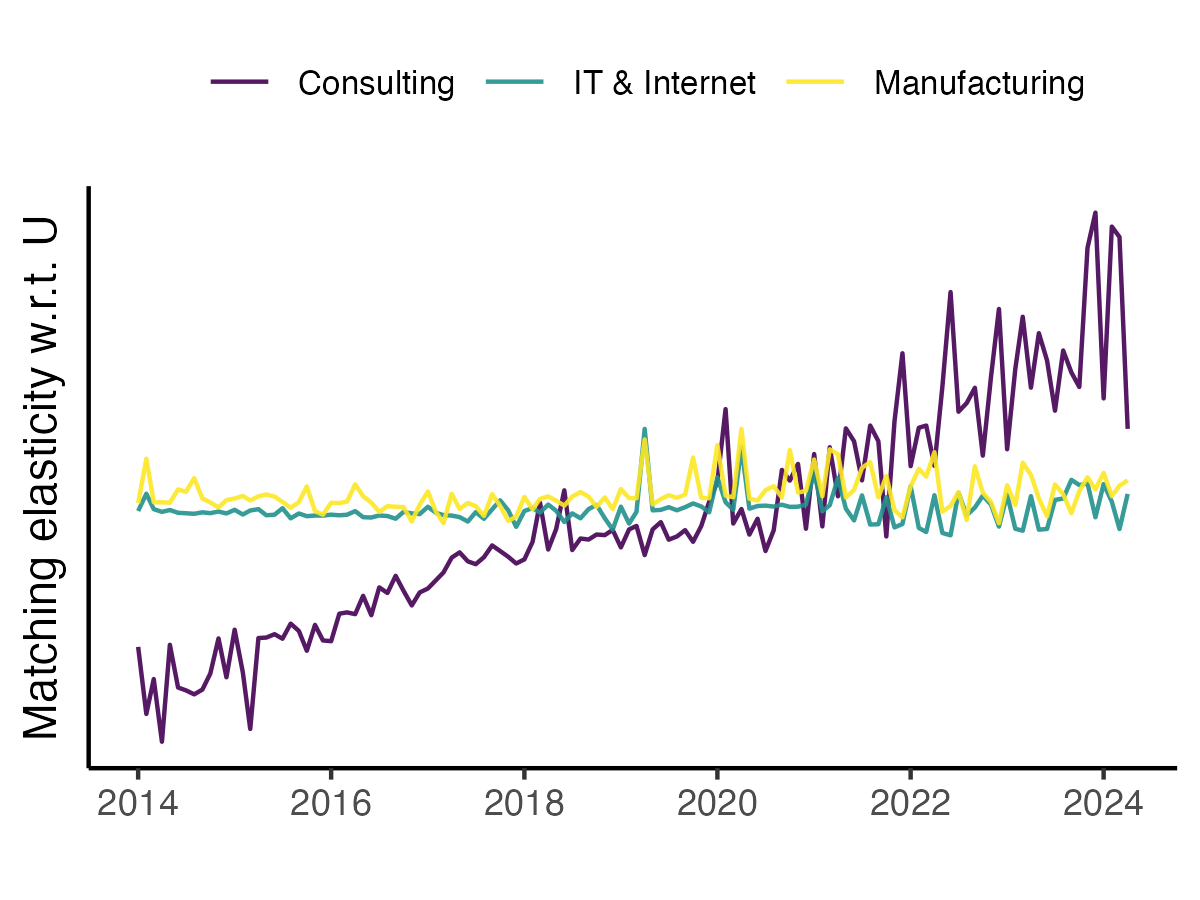}}
  \subfloat[Matching Elasticity ($\frac{d\ln m}{d\ln V}$)]{\includegraphics[width = 0.37\textwidth]
  {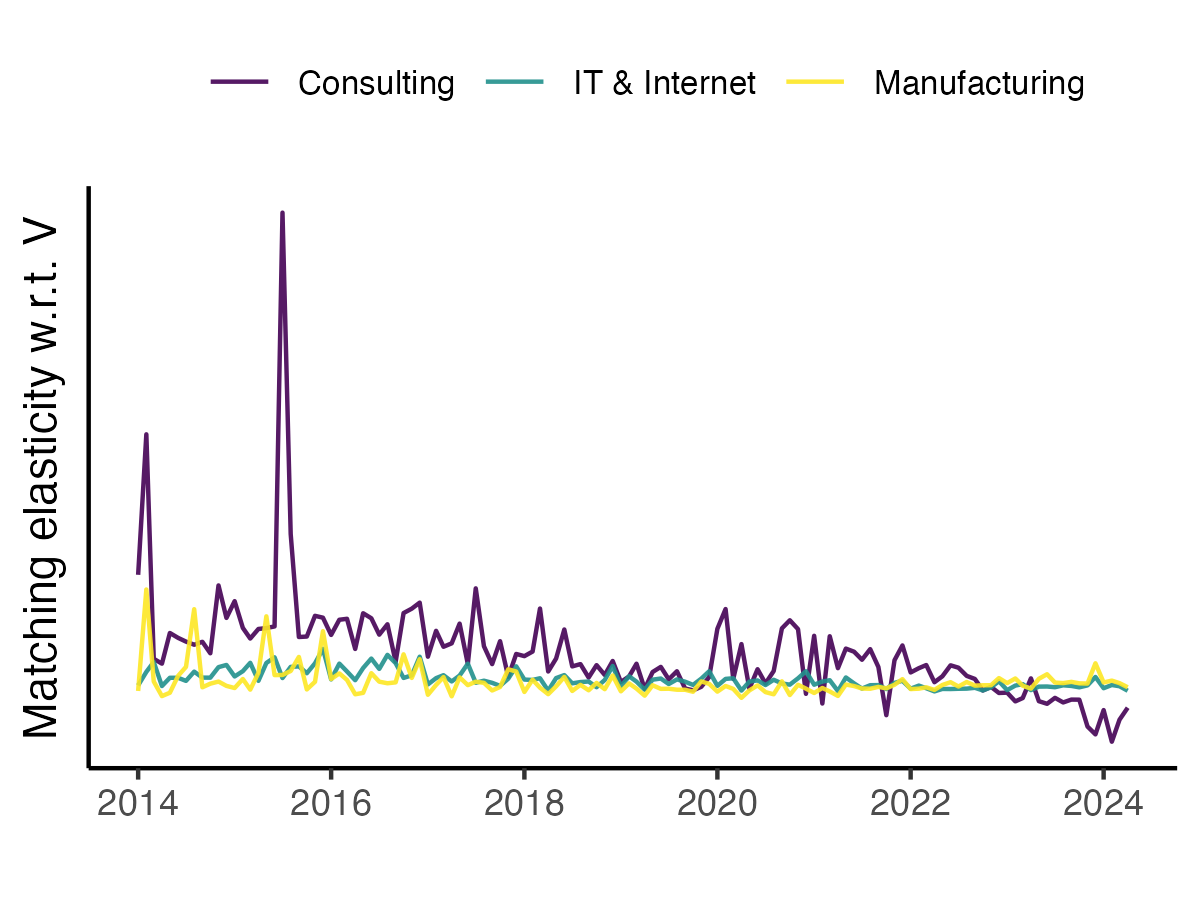}}
  \caption{Industry-level results on Platform 2014-2024}
  \label{fg:matching_efficiency_month_aggregate_each_industry} 
  \end{center}
  \footnotesize
  Note: For confidentiality reasons, the y-axis levels are masked.
\end{figure} 

Figure \ref{fg:matching_efficiency_month_aggregate_each_industry}  illustrates labor market dynamics across three sectors—Consulting, IT and Internet, and Manufacturing—on a private job platform from 2014 to 2024. These categories represent high-skill full-time employed workers, though they differ from those used by Hello Work. For confidentiality reasons, the precise numbers of users, vacancies, hires, and the y-axis levels are not disclosed, but the trends provide valuable insights into sector-specific patterns.

Panel (a) shows that labor market tightness ($V/U$) has generally increased across all three sectors from 2014 to 2024. Consulting and IT \& Internet sectors exhibit higher and more volatile tightness ratios, particularly after 2020, reflecting more job openings per user. In contrast, the Manufacturing sector shows a lower and more stable tightness ratio, indicating fewer vacancies relative to job seekers in this field.

Panel (b) depicts matching efficiency, with the IT and Internet sector exhibiting the highest variability and levels, especially after 2020. Panels (c) and (d) illustrate the matching elasticities with respect to users and vacancies, respectively. The Consulting sector shows negative elasticity with respect to users early on, likely due to a small user base, but this increases sharply over time. Meanwhile, elasticity with respect to vacancies slightly decreases. IT and Internet and Manufacturing sectors show more stable elasticity patterns. The overall trends highlight significant industry-level heterogeneity, with Consulting experiencing the most dynamic growth, while IT and Internet and Manufacturing follow more stable but slower trends.

\subsection{Policy Implication}

Japan presents a particularly interesting case for studying matching efficiency due to its unique labor market institutions. Despite historically low job mobility and strong norms around lifetime employment, recent policy reforms (e.g., the 2018 Work Style Reform Laws) have actively encouraged more flexible employment transitions, especially for mid- and high-skilled workers. Understanding how these institutional shifts have affected actual matching efficiency over time is crucial for evaluating the effectiveness of such reforms, mainly affecting on-the-job-search workers. Moreover, Japan offers a rare setting where detailed aggregate data from both a nationwide public employment agency and a large private job platform without mergers are available over a decade. 

Although Hello Work and BizReach serve different populations and operate under distinct matching mechanisms (application-based vs. scouting-based), this contrast is precisely what makes the comparison informative. By examining both platforms through a unified matching function framework, we can quantify how different institutional settings and search technologies affect labor market efficiency. The comparison sheds light on the relative strengths and weaknesses of traditional government-run matching systems versus emerging private platforms in adapting to a changing labor market. This is particularly relevant for policymakers contemplating reforms or partnerships involving hybrid public–private matching infrastructures.

Our analysis does not aim to equate the two markets, but rather to compare system-level efficiency patterns under a consistent empirical framework. This comparison offers insights into which institutional designs and targeting strategies are more effective given different worker attributes.
Moreover, the findings about the higher responsiveness and volatility of the private platform underscore how market design (e.g., proactive scouting) and targeting (e.g., high-skill professionals) interact with search intensity and match outcomes. These results are particularly timely given recent efforts to reform Japan’s public employment services. For example, Hello Work has launched pilot programs integrating AI-based recommendation systems to improve job matching accuracy, signaling a recognition that conventional application-based approaches may underperform for certain segments or in rapidly changing market conditions. This underscores the policy relevance of comparative studies on institutional matching efficiency.\footnote{Ministry of Health, Labour and Welfare, Employment Security Bureau (2025), Looking Ahead: AI Utilization in Hello Work Employment Services, April 22, 2025. Available at: \url{https://www.mhlw.go.jp/content/11601100/001478507.pdf}.}

The recent report that the hiring rate through Hello Work has fallen to a record low of 11.6\% highlights critical structural challenges in Japan’s public employment services.\footnote{Nihon Keizai Shimbun (2025), “Hello Work Job Placement Rate Falls to Record Low of 11.6\%—Rising Reliance on Private Services,” May 20, 2025. Available at: \url{https://www.nikkei.com/article/DGKKZO88780940Z10C25A5EP0000/}.} Our findings, which reveal higher responsiveness and elasticity in the private platform compared to the public one, suggest that different market aspects and user targeting strategies lead to divergent performance. This raises important policy questions: Should the segmentation between public services for lower-skilled unemployed workers and private platforms for higher-skilled employed individuals be maintained, or is there merit in promoting a more integrated model? One potential direction is to allow private platforms to assume responsibility for specific functions of Hello Work—such as digital matching for mid-career or high-skill segments—while maintaining the public agency’s role in providing universal access. Alternatively, reforming Hello Work itself through technological enhancements, including its ongoing AI-based recommendation pilot programs, may offer a path to improving matching outcomes without structural overhaul. These questions are central to ongoing debates about privatization, outsourcing, and the optimal configuration of employment intermediation in aging and skill-diverse labor markets like Japan’s.

\section{Conclusion}
This paper uses proprietary data from BizReach, an online job scouting platform in Japan, spanning from 2014 to 2024, to estimate the matching function for high-skill employed workers in a private on-the-job search platform. The results are compared to a public off-the-job search platform, specifically targeting unemployed workers seeking full-time jobs. Findings suggest that matching efficiency on the private platform is more volatile but generally higher than on the public platform, highlighting the increasing reliance on private platforms for high-skill job searches. Furthermore, the private platform demonstrates a higher matching elasticity with respect to unemployment (between 0.6 and 0.8) compared to the public platform, while the elasticity with respect to vacancies is similar (between 0.8 and 1.1). This indicates a more balanced responsiveness to changes in both users and vacancies on the private platform, compared to the more stable but less dynamic public platform, Hello Work.

Additionally, industry-level heterogeneity is evident across both platforms, reflecting differing labor market dynamics by sector. However, while this paper provides key insights into the matching function for on-the-job searches, the analysis may not fully represent the broader on-the-job search labor market, particularly for non-high-skill workers in Japan. Moreover, the standard assumption of homogeneity among workers and vacancies may overlook important nuances. Future research should focus on expanding this analysis to other private platforms and exploring individual-level behavior, as discussed in studies like \cite{kambayashi2025decomposing} and \cite{roussille2023bidding}, to provide a more comprehensive understanding of labor market dynamics.
\newpage
\bibliographystyle{ecca}
\bibliography{matching_function}

\appendix
\section{Appendix}\label{sec:appendix_cobb_douglas}

Table \ref{tb:estimation_results_cobb_douglas} presents country–month level estimates using Hello Work data to illustrate model comparisons. In Panel (a), the country–month level Cobb–Douglas estimates appear implausible: the constant term suggests negative matching efficiency, and the elasticities sum to well above unity, violating the constant-returns-to-scale condition. These issues likely stem from aggregation at the national level, which masks regional and occupational heterogeneity and inflates the responsiveness of hires to unemployment and vacancies. In addition, the small number of observations reduce stability. By contrast, Panel (b) shows that estimation at finer levels of aggregation—such as prefecture–month or occupation–month—yields more plausible results with fixed effects, with reasonable elasticities summing closer to one. Taken together, these patterns suggest that the implausible results should not be interpreted as evidence against the constant returns to scale and positive efficiency per se, but rather highlight that the estimates are sensitive to the level of aggregation and the sample period used.

While the results for BizReach are confidential and cannot be reported, the estimates are more reasonable than the Hello Work results. Moreover, \cite{otani2024nonparametric} demonstrates through Monte Carlo simulations that the nonparametric approach yields lower bias and Root Mean Squared Error (RMSE) and is robust to functional form misspecification, further supporting the usefulness of this framework.

Table \ref{tb:estimation_results_cobb_douglas} compares model performance across datasets. In the case of BizReach, where matching efficiency is time-varying, the nonparametric approach provides superior performance in terms of RMSE and Mean Absolute Error (MAE). For Hello Work, however, where matching efficiency appears relatively stable over 2014–2024, the constant-efficiency OLS specification delivers a better fit by suppressing fluctuations. Nevertheless, such stability is itself revealed only when allowing for nonparametric estimation, underscoring the continued value of the nonparametric approach.

\begin{table}[!htbp]
  \begin{center}
      \caption{Estimation Results Using Cobb-Douglas Matching Function: Hello Work Data}
      \label{tb:estimation_results_cobb_douglas}       
      \subfloat[Country-Month Level]{\input{figuretable/matching_function_project/estimation_results_cobb_douglas}}
      \subfloat[Prefecture-Month or Occupation-Month Level]{\input{figuretable/matching_function_project/from_hellowork_paper/estimation_results_cobb_douglas}}

  \end{center}\footnotesize
  \textit{Note}: FE: fixed effects. In Panel (a), I use the Report on Employment Service (\textit{Shokugyo Antei Gyomu Tokei}) for country-month-level aggregate data from January 2014 to April 2024 to examine trends in matching unemployed full-time workers with vacancies via Japan's public employment platform, Hello Work. In Panel (b), I extract results from \cite{otani2024nonparametric} using the prefecture-month-level and occupation-month-level data from January 2012 to March 2023.
\end{table}

\begin{table}[!htbp]
  \begin{center}
      \caption{Model Performance Comparison}
      \label{tb:rmse_mae_bizreach_data}       
      \subfloat[BizReach Data]{\input{figuretable/matching_function_project/rmse_mae_bizreach_data}}\\
      \subfloat[Hello Work Data]{\input{figuretable/matching_function_project/rmse_mae_hellowork_data}}
      
  \end{center}\footnotesize
  \textit{Note}: RMSE = Root Mean Squared Error, MAE = Mean Absolute Error.
\end{table}

\end{document}

%% file: figuretable/matching_function_project/estimation_results_cobb_douglas.tex
\begin{tabular}[t]{lc}
\toprule
  & (1)\\
\midrule
Dependent var & log(H)\\
Constant & -56.673\\
 & (4.661)\\
log(U) & 2.688\\
 & (0.180)\\
log(V) & 2.123\\
 & (0.202)\\
\midrule
Num.Obs. & 124\\
R2 & 0.661\\
R2 Adj. & 0.656\\
RMSE & 0.16\\
\bottomrule
\end{tabular}

%% file: figuretable/matching_function_project/from_hellowork_paper/estimation_results_cobb_douglas.tex
\begin{tabular}[t]{lcc}
\toprule
  & (1) & (2)\\
\midrule
Dependent var & log(H) & log(H)\\
log(U) & 0.266 & 0.510\\
 & (0.019) & (0.013)\\
log(V) & 0.319 & 0.546\\
 & (0.014) & (0.010)\\
\midrule
FE & pref, month & occupation, month\\
Num.Obs. & 6392 & 8908\\
R2 & 0.987 & 0.992\\
R2 Adj. & 0.987 & 0.992\\
\bottomrule
\end{tabular}

%% file: figuretable/matching_function_project/rmse_mae_bizreach_data.tex
\begin{tabular}[t]{lrr}
\toprule
 & Cobb Douglas & Nonparametric\\
RMSE & 0.1834 & 0.1821\\
MAE & 0.1387 & 0.1237\\
\bottomrule
\end{tabular}

%% file: figuretable/matching_function_project/rmse_mae_hellowork_data.tex
\begin{tabular}[t]{lrr}
\toprule
 & Cobb Douglas & Nonparametric\\
RMSE & 0.1639 & 0.2006\\
MAE & 0.1359 & 0.1579\\
\bottomrule
\end{tabular}